\documentclass{sig-alternate}
\usepackage{graphicx}
\usepackage{colortbl}
\usepackage{subfigure}
\usepackage{multirow}
\usepackage{fancyvrb,relsize}

\begin{document}
\conferenceinfo{HT'10,} {June 13--16, 2010, Toronto, Ontario, Canada..} 
\CopyrightYear{2010}
\crdata{978-1-4503-0041-4/10/06}
\clubpenalty=10000
\widowpenalty = 10000
\title{Is This a Good Title?}
\numberofauthors{3}
\author{
\alignauthor
Martin Klein\\
       \affaddr{Department of Computer Science}\\
       \affaddr{Old Dominion University}\\
       \affaddr{Norfolk, VA,  23529}\\
       \email{mklein@cs.odu.edu}
\alignauthor
Jeffery Shipman\\
       \affaddr{Department of Computer Science}\\
       \affaddr{Old Dominion University}\\
       \affaddr{Norfolk, VA,  23529}\\
       \email{jshipman@cs.odu.edu}
\alignauthor
Michael L. Nelson\\
       \affaddr{Department of Computer Science}\\
       \affaddr{Old Dominion University}\\
       \affaddr{Norfolk, VA,  23529}\\
       \email{mln@cs.odu.edu}
}
\maketitle
\begin{abstract}
Missing web pages, URIs that return the $404$ ``Page Not Found'' error or the HTTP response code $200$ but
dereference unexpected content, are ubiquitous in today's browsing experience.
We use Internet search engines to relocate such missing pages and provide means that help automate the rediscovery process.
We propose querying web pages' titles against search engines. We investigate the retrieval performance of 
titles and compare them to lexical signatures which are derived from the pages' content.
Since titles naturally represent the content of a document they intuitively change over time.
We measure the edit distance between current titles and titles of copies of the same pages obtained from the Internet
Archive and display their evolution.
We further investigate the correlation between title changes and content modifications of a web page over time.
Lastly we provide a predictive model for the quality of any given web page title in terms of its 
discovery performance.
%
Our results show that titles return more than $60\%$ URIs top ranked and further relevant content returned in the
top $10$ results.
We show that titles decay slowly 
but are far more stable than the pages' content.
We further distill stop titles than can help identify insufficiently performing search engine queries.
\end{abstract}

\category{H.3.3}{Information Storage and Retrieval}Information Search and Retrieval

\terms{Measurement, Performance, Design}
\keywords{Web Page Titles, Web Page Discovery, Digital Preservation}

\section{Introduction} \label{sec:intro}
\begin{figure}[t]
\begin{center}
 \subfigure[Original URI, new (unrelated) Content at \texttt{http://www.ht06.org/}]{\label{fig:ht06_1}\includegraphics[scale=0.23]{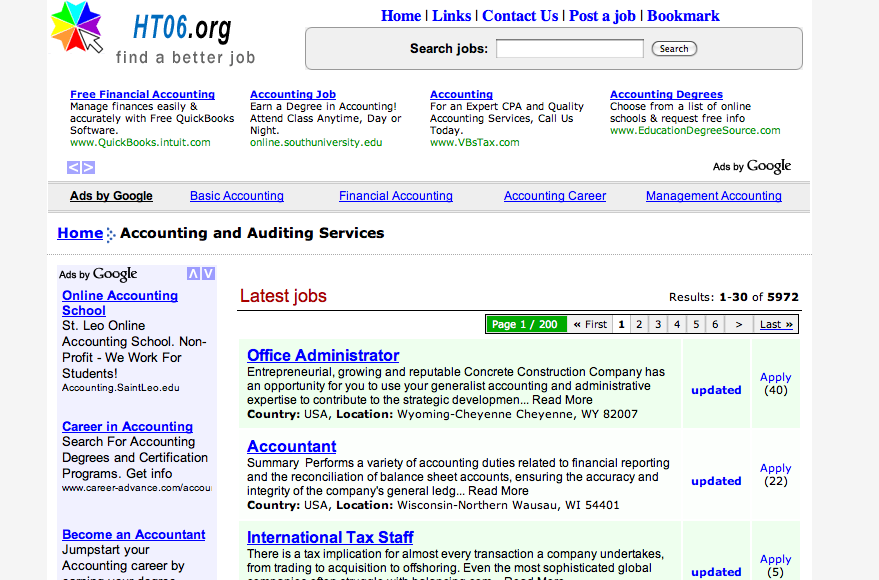}}
 \vspace{.1in}
 \subfigure[Original Content, new URI \texttt{http://hypertext.expositus.com/}]{\label{fig:ht06_2}\includegraphics[scale=0.2]{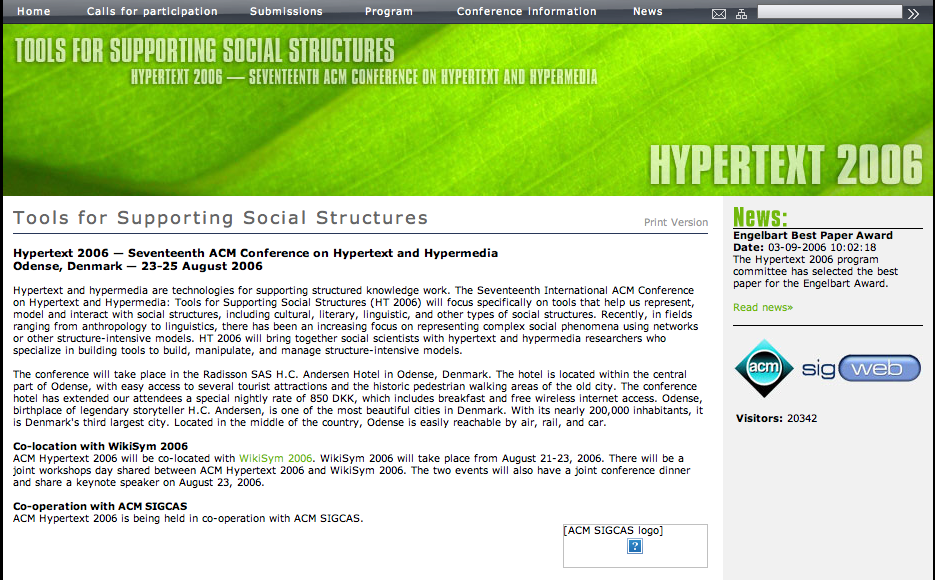}}
 \caption{The Content of the Website for the Conference Hypertext $2006$ has Moved over Time}
 \label{fig:ht06}
\end{center}
\end{figure}

Inaccessible web pages and ``404 Page Not Found'' responses are part of the web browsing experience.
Despite guidance for how to create ``Cool URIs'' that do not change \cite{berners-lee:cool} there are many reasons why URIs
or even entire websites break \cite{marshall:archiving_strategies}.
Since web users frequently re-visit web pages \cite{adar:revisitation_patterns} a $404$ response constitutes a detriment
to their browsing experience.
However, we claim that information on the web is rarely completely lost, it is just missing.
In whole or in part, content is often just moving from one URI to another. 
\begin{figure}[ht]
 \centering
 \includegraphics[scale=0.23]{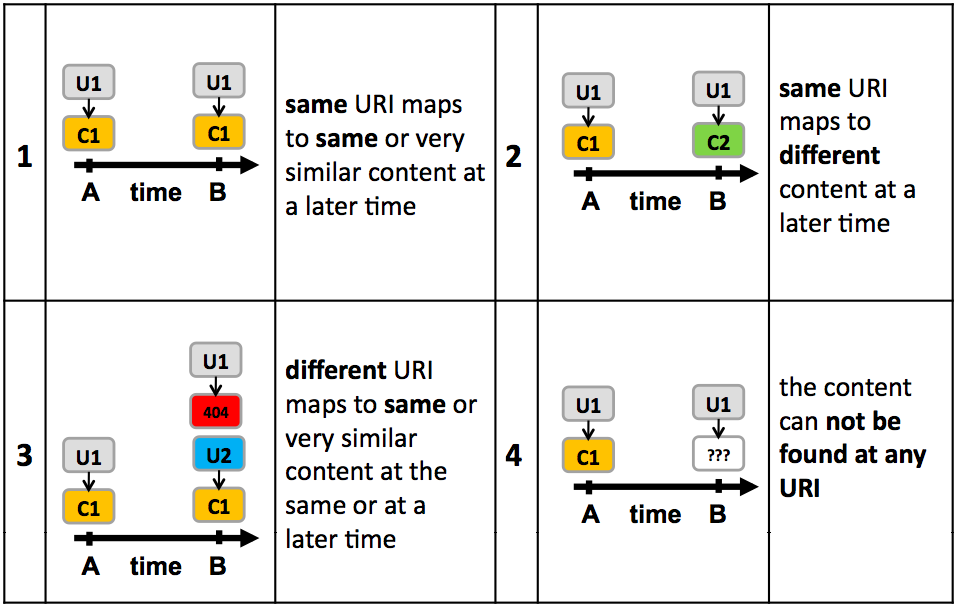}
 \caption{The URI Content Mapping Problem}
 \label{fig:content_mapping}
\end{figure}
Figure \ref{fig:content_mapping} graphically explains this URI content mapping problem showing four scenarios with
URIs ($U$) mapping to the same and to different content ($C$) over time.
Furthermore Figure \ref{fig:ht06} shows an example of a web page whose content has moved within a three year period.
Figure \ref{fig:ht06_1} shows the content of the original URI of the Hypertext $2006$ conference as displayed in 8/2009.
The original URI clearly does not hold conference related content anymore. Our suspicion is that the website administrators
did not renew the domain registration and therefore enabling someone else to take over. However, the content is not lost. 
The title of the original web page was \textit{ACM Hypertext 2006}. Querying it against today's search engines results in
discovering the content at its new URI. Yahoo and Bing return the new page top ranked and Google returns it ranked fourth.
Figure \ref{fig:ht06_2} shows the content which is now available at a new URI.

It is our intuition that major search engines like Google, Yahoo and MSN Live (now Bing), as members of what we call
the Web Infrastructure (WI), likely have crawled the content and possibly even stored a copy in their cache.
Therefore the content is not lost, it ``just'' needs to be rediscovered.
The WI, explored in detail in \cite{jatowt:browser,mccown:thesis,nelson:web-infrastructure}, also includes 
non-profit archives such as the Internet Archive (IA) or the European Archive as well as large-scale academic
digital data preservation projects e.g., CiteSeer and NSDL.

It is our goal to utilize the WI for digital preservation and in particular for the rediscovery of missing web pages. 
Therefore we need to explore the notion of the ``aboutness'' of the missing pages.
Lexical signatures (LSs) haven been shown to be suitable for this purpose
\cite{klein:ls,klein:methods_rediscover,park:ls-tois,phelps:hyperlinks}
but they are expensive to generate since the inverse document frequency (IDF) value needs to be acquired for each candidate term 
for example by querying search engines. In the worst case the cost is one query for each term.
In this paper we investigate web pages' titles as they intuitively describe the ``aboutness'' of a web page.
Other measures such as hashes \cite{charikar:simhash} and shingles \cite{broder:syntactic_clustering} have been introduced to
capture and compare document's content but since we are utilizing search engines we need to leverage textual queries since
that is the only format public search engine interfaces process.
We can obtain titles of missing pages from search engine caches and the IA which means it ideally requires only one query per title.
However, it implies that our method is only applicable for web pages (html files) with titles.
%
For the sake of simplicity we assume all titles to be in a search engine's tolerance range for a search query. 
We are aware, however, that we obtain these web pages' titles from third party institutions and the title text is often created by humans
which both bares a certain risk regarding its correctness.

We focus on the following four aspects of web pages' titles: 
%
1) 
their retrieval performance in terms of the rank of the URI of interest in the result set and the 
 degree of similarity between the content of the top $10$ results and the content of the URI of interest;
2) 
their evolution over time measured in edit distance;
3) 
the correlation between the change of titles and the change of their page content
and 4)
a predictive model of a well performing title. 
%
%

%
%
\begin{table*}[ht]
\centering
  \begin{tabular}{|c||c|c|} \hline
        &\textbf{www.nicnichols.com}&\textbf{www.verticalradio.org}\\ \hline \hline
        \multirow{0}{*}{\textbf{LS}}&Nichichols Nichols Nic Stuff Shoot & Vertical Radio God Knmi Station \\ \hline
        \multirow{2}{*}{\textbf{Title}}&Documentary Toy Camera Photography of Nic & Home \\
	&Nichols: Holga, Lomo and other Lo-Fi Cameras!& \\ \hline
  \end{tabular}
  \caption{Lexical Signatures and Titles Obtained from \texttt{www.nicnichols.com} and \texttt{www.verticalradio.org}}
 \label{tab:ls_title_example}
\end{table*}
%

The following two examples further demonstrate the motivation of this research.
Table \ref{tab:ls_title_example} shows the LS and title of two web pages \texttt{www.nicnichols.com} and \texttt{www.verticalradio.org}.
The first page is about the photographer Nic Nichols and the second about a Christian radio station in New Mexico.
Both LSs look promising and indeed return the page as the top ranked result when queried against the Yahoo search engine. The title of
the first page also returns the URI top ranked by the title of the radio station's web page clearly is not sufficient as a search engine
query. Even though Yahoo has indexed the URI with the term \textit{HOME} querying it returns more than one billion results (through the BOSS
API\footnote{\texttt{http://developer.yahoo.com/search/boss/}}) and the URI is not returned within the top $100$ results.

The contribution of this paper is a discussion of the discovery performance of web pages' titles compared to LSs,
the relevancy of the returned results, the evolution of titles over time in comparison to the change of web pages' content
and the introduction of a prediction model to assess the title's retrieval potential in real time.
\section{Related Work} \label{sec:relwork}
\subsection{Missing Web Pages}
Missing web pages are a pervasive part of the web experience.
The lack of link integrity on the web has been addressed by numerous researchers
\cite{ashman:document_addressing,ashman:missing404,davis:ht_link_integrity,davis:referential_integrity}.
In $1997$ Brewster Kahle published an article focused on preservation of Internet resources claiming that the expected lifetime of a web page is $44$
days \cite{kahle:preserving}.
A different study of web page availability performed by Koehler \cite{koehler:web-page-change} shows the random test collection of URIs
eventually reached a ``steady state'' after approximately $67$\% of the URIs were lost over a $4$-year period.
Koehler estimated the half-life of a random web page is approximately two years.
Lawrence et al. \cite{lawrence:persistence} found in $2000$ that between $23$ and $53\%$ of all URIs occurring in computer science
related papers authored between $1994$ and $1999$ were invalid. By conducting a multi level and partially manual search on the Internet,
they were able to reduce the number of inaccessible URIs to $3$\%. This confirms our intuition that information is rarely lost, it is just moved.
This intuition is also supported by Baeza-Yates et al. \cite{baeza-yates:genealogical_trees} who show that a significant portion of the web
is created based on already existing content.

Spinellis \cite{spinellis:decay} conducted a study investigating the accessibility of URIs occurring in papers published in
Communications of the ACM and IEEE Computer Society. He found that $28$\% of all URIs were unavailable after five years and $41\%$ after seven years.
He also found that in $60\%$ of the cases where URIs where not accessible, a $404$ error was returned. He estimated the half-life of an URI
in such a paper to be four years from the publication date.
Dellavalle et al. \cite{dellavalle:going} examined Internet references in articles published in journals with a high impact factor given by the
Institute for Scientific Information (ISI).
They found that Internet references occur frequently (in $30\%$ of all articles) and are often inaccessible within months after publication
in the highest impact (top $1\%$) scientific and medical journals. They discovered that the percentage of inactive references (references that return
an error message) increased over time from $3.8\%$ after $3$ month to $10\%$ after $15$ month up to $13\%$ after $27$ month.
The majority of inactive references they found were in the \emph{.com} domain ($46\%$) and the fewest in the \emph{.org} domain ($5\%$).
By manually browsing the IA they were able to recover information for about $50\%$ of all inactive references.
\subsection{Near-Duplicate Web Pages}
When thinking about discovering missing web pages we consider research in the field of (near-)duplicate web page detection as relevant.
Even though most introduced techniques are intended for optimizing the process of web crawling (by identifying and hence omitting duplicates)
they can still be of use for web page preservation.
Broder et al. \cite{broder:syntactic_clustering} introduced shingles as a technique to estimate the syntactic similarity between web pages.
Web pages can be clustered based on their shingle values which for example can be applied for what the authors propose as a ``Lost and Found''
service for web pages. 
Charikar \cite{charikar:simhash} introduced another technique that became very popular - a hashing function that changes relative to the
changes of the input set. That means entire web pages or subset of pages can be compared by their hash values.
Monika Henzinger \cite{henzinger:finding_near-duplicates} 
found that both techniques perform well on identifying
(near-)duplicates on different sites but not on the same site. She proposes a combination of both methods to overcome that weakness.

Fetterly et al. \cite{fetterly:evolution_of_clusters} created clusters of near-duplicate web pages.
They found that about $28\%$ of their pages were duplicates and $22\%$ were virtually identical.
Their results also support the intuition that a lost page often can be restored by finding other pages on the web.

The work done by Brin et al. \cite{brin:copy_detection} is also related since they introduced further methods to detect
copied documents by comparing ``chunks'' of the documents. In \cite{brin:near_neighbor} Brin transforms text into a metric space
and computes the document similarity based on the distances in the metric space.
%

%
Adar et al. \cite{adar:the_web_changes_everything} conducted a study exploring changes of web pages. They analyzed change on the content
level, the term level and structural level of $55.000$ web pages over a five week period. Roughly $65\%$ of their pages showed
some change while the degree of change depends on domain and structure of the page. They identify page specific ephemeral
vocabulary as well as terms with high staying power, both potentially useful to determine the page's ``aboutness''.
While the authors find that various structural elements of web pages change with different rates they do not specifically mention titles.
\subsection{Search Engine Queries}
The work done by Henzinger et al. \cite{henzinger:query-free} is related in the sense that they tried to determine the ``aboutness'' of news documentations.
They provide the user with web pages related to TV news broadcasts using a $2$-term summary which can be thought of as a LS.
This summary is extracted from closed captions of the broadcast and various algorithms are used to compute the scores determining the most relevant terms.
The terms are used to query a news search engine while the results must contain all of the query terms.
The authors found that $1$-term queries return results that are too vague and $3$-term queries return too often zero results.
Thus they focus on creating $2$-term queries.
%

Bharat and Broder \cite{bharat:size_and_overlap_of_ses} investigated the size and overlap of search engine indexes. They tried to
discover the same page in several different search engine indexes. Their technique was generally based in randomly sampling URIs from
one index and checking whether they exist in another. They introduce the notion of ``strong queries'', a set of salient keywords
representing the randomly sampled URI and used as the query against the other indexes, hoping it would return the URI.
They simply used the $n$ terms from the document that least frequently occurred in their entire data set and formed a conjunctive query
with the ``AND'' operator.

He and Ounis' work on query performance prediction \cite{he:inferring_query_performance} is based on the TREC dataset. They measured retrieval performance of
queries in terms of average precision (AP) and found that the AP values depend heavily on the type of the query.
They further found that what they call \textit{simplified clarity score (SCS)} has the strongest correlation with AP for title queries (using the title of
the TREC topics).
SCS depends on the actual query length but also on global knowledge about the corpus such as document frequency and total number of tokens in the corpus.
\subsection{Lexical Signatures of Web Pages}
Phelps and Wilensky \cite{phelps:hyperlinks} introduced the term lexical signature and first proposed
their use for finding content that had moved from one URI to another. Their claim was ``robust hyperlinks cost just 5 words each'' and their
preliminary tests confirmed this. The LS length of $5$ terms however was chosen somewhat arbitrarily. Phelps and Wilensky proposed ``robust hyperlinks'',
an URI with a LS appended as an argument. They conjectured that if an URI would return a $404$ error, the browser would submit its appended LS 
to a search engine in order to find the relocated copy.

Park et al. \cite{park:ls-tois} expanded on the work of Phelps and Wilensky, studying the performance of $9$ different LS generation algorithms
(and retaining the $5$-term precedent). The performance of the algorithms depended on the intention of the search.
Algorithms weighted for term frequency (TF; ``how often does this word appear in this document?'') were better at finding related pages, but the exact page
would not always be in the top N results.
Algorithms weighted for inverse document frequency (IDF; ``in how many documents of the entire corpus does this word appear?'') were better at finding the
exact page but were susceptible to small changes in the document (e.g., when a misspelling is fixed).
\section{Experiment Setup} \label{sec:ex_setup}
We are not aware of a data corpus providing missing web pages.
We therefore generate a dataset of URIs taken from the live web and ``pretend'' they are missing.
We know they are indexed by search engines so by querying the right terms, we will be able to retrieve them in the result set.
\subsection{Data Gathering}
As shown in \cite{henzinger:url_sampling,rusmevichientong:sampling_pages,theall:methodologies}, finding a small sample set of URIs representative
for the entire Internet is not trivial. 
Rather than attempt to get an unbiased sample, we randomly sampled $20,000$ URIs from the Open Directory Project \texttt{dmoz.org}.
We are aware of the implicit bias of this selection. Given the great popularity of DMOZ it is reasonable to assume that search engines have
indexed these pages.
Furthermore this bias applies to the IA index since the best effort crawling approach the IA is taking presumably
focuses on popular hubs and pages first. Our assumption also is that if a web site administrator invests the time and effort to register the site
with DMOZ it is likely well structured and crawler friendly.

Starting with our $20,000$ URIs we applied the two filters that were used in \cite{klein:ls,park:ls-tois}.
We dismissed all pages containing less than $50$ terms and applied a very restrictive off-the-shelf English language
filter (the Perl package \textit{Lingua::Identify} available through CPAN). This process shrank our corpus down to $6,875$ pages.
Table \ref{tab:url_stats} shows the top level domains of the originally sampled URIs and of the final filtered dataset.
We downloaded the content of all $6,875$ pages in July and August of $2009$ and excluded all HTML elements.
\begin{table}
 \centering
  \begin{tabular}{|c||c|c|c|c|c|} \hline
	&\texttt{.com}&\texttt{.org}&\texttt{.net}&\texttt{.edu}&\textbf{Sum} \\ \hline \hline 
	\textbf{Original}&15289&2755&1459&497&20000 \\ \hline 
	\textbf{Filtered}&4863&1327&369&316&6875 \\ \hline 
  \end{tabular}
  \caption{Sample Set URI Statistics}
 \label{tab:url_stats}
\end{table}
\subsection{Title Extraction and Copies from the\\Internet Archive}
Titles of web pages are commonplace. Only $0.6\%$ of our web pages (a total of $41$) did not have a title.
We extracted all titles by simply parsing the page and extract everything between the HTML tags \texttt{<title></title>}.

In order to investigate the temporal aspect of the title evolution we queried the URIs against the Internet Archive (IA). 
The IA and its crawler is not in competition with search engines. It rather is a best effort approach and all copied pages remain in 
a ``quarantine period'' for six to up to 12 month before they become accessible through the IA interface.
Out of our $6,875$ pages the IA provided copies for $6,093$ URIs. We downloaded all available copies (more than $500,000$) and extracted the
pages' content and titles.
\subsection{LS Generation of Web Pages}
For the purpose of comparison we also generated LSs of our $6,875$ pages.
We have shown in our earlier research \cite{klein:ls} that LSs containing $5$- and $7$-terms perform best and hence 
we generated two LSs per URI following the TF-IDF scheme which was also applied in \cite{park:ls-tois}.
We used the ``screen scraping'' approach and queried the Yahoo BOSS
API to determine document frequency values for all terms and used numbers published by the website \texttt{www.worldwidewebsize.com}
in October $2009$ to estimate the size of the Yahoo index. We have shown in \cite{klein:idf} that this approach is feasible and performs very well
compared to other methods.
\section{Title and LS Performance}
We took all LSs and titles and queried them against the Yahoo BOSS search API.
We are aware of APIs available from Google and MSN but they either impose restrictions in terms of number of queries per day or have
been shown to perform not quite as good as the BOSS API \cite{klein:cross_se_ls,klein:methods_rediscover}. 
In terms or retrieval performance we distinguish between four cases for an URI to be discovered:
\begin{itemize}
\item top ranked
\item ranked within the top $10$ but not top
\item ranked within the top $100$ but beyond the top $10$
\item ranked somewhere beyond the top $100$ (undiscovered).
\end{itemize}
In the last scenario we consider the URI as undiscovered since the majority of the search engine users do not browse through the result set past the top
$100$ results.
%
%
With these scenarios we evaluate our results as success at $1$, $10$ and $100$. Success is defined as a binary value, as the target either occurs in
the subset (top result, top $10$, top $100$) of the entire result set or it does not. For simplicity we dismiss all query string parameters that a URI
may contain and match only the remaining part of the URI.

\begin{figure}[t]
\centering
 \subfigure[$5$- and $7$-Term LS Performance]{\label{fig:ls_retrieval}\includegraphics[scale=0.33]{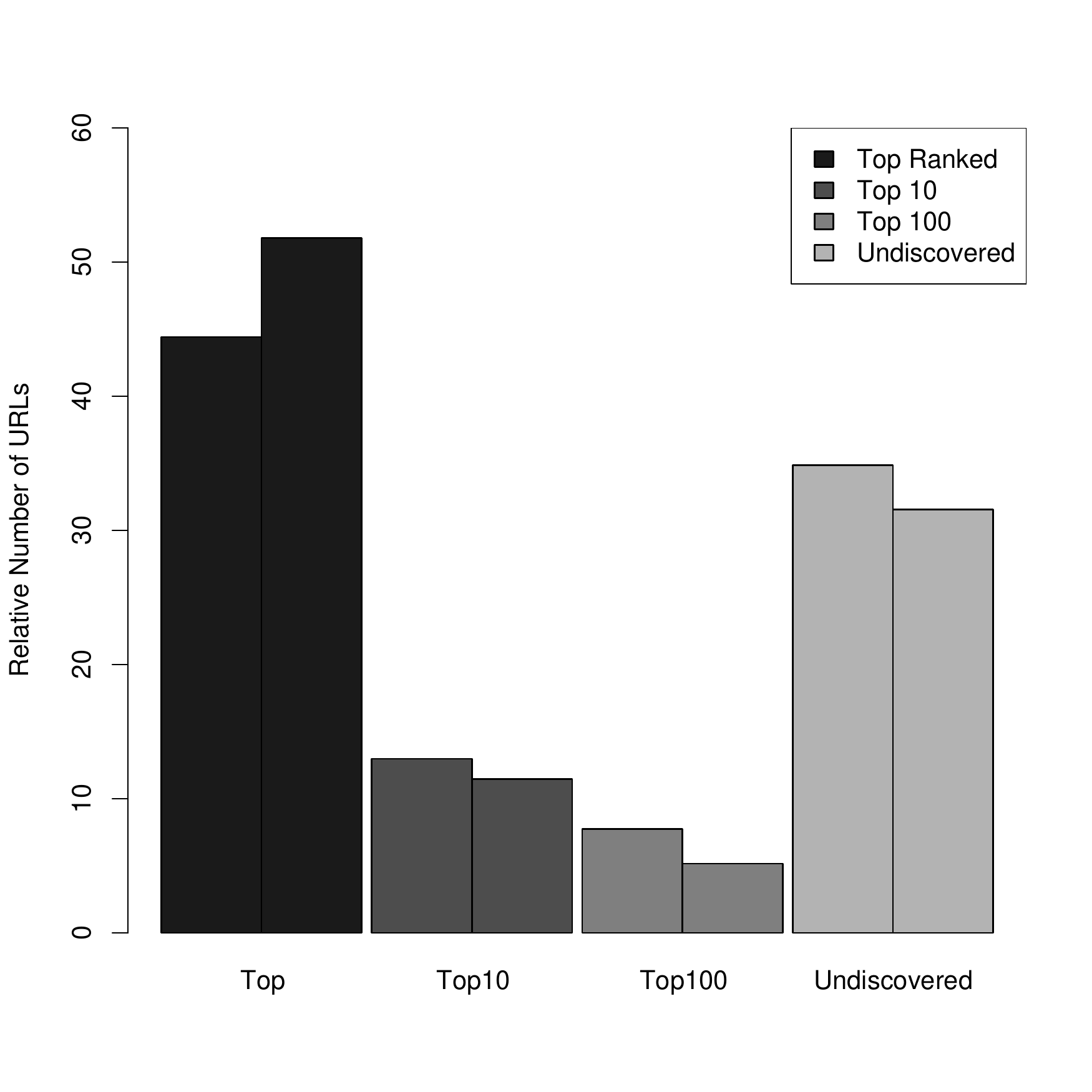}}
 \subfigure[Title Performance]{\label{fig:title_retrieval}\includegraphics[scale=0.33]{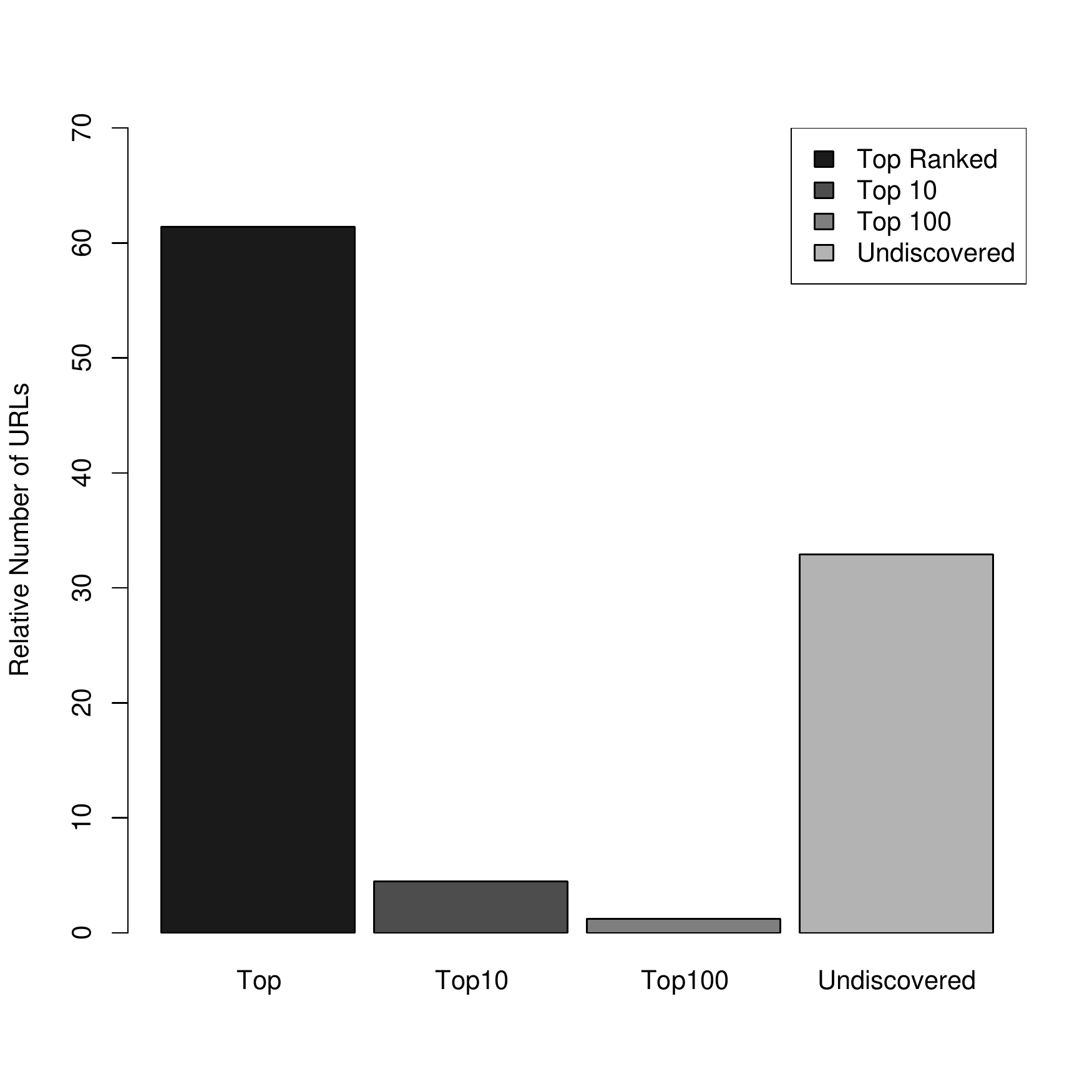}}
 \caption{Retrieval Performance of LSs and Web Pages' Titles}
 \label{fig:retrieval}
\end{figure}
Figure \ref{fig:ls_retrieval} shows the relative number of retrieved URIs when querying the LSs of the pages distinguished by number of terms.
The four tuples of bars represent the four scenarios mentioned above and each tuple contains one bar showing the results for $5$- and one 
for $7$-term LSs. We see more than $50\%$ of the URIs returning top ranked and about $30\%$ remain undiscovered with $7$-term LSs.
Our numbers for $5$-term LSs are slightly lower but still comparable with the findings by Park et al \cite{park:ls-tois}.
Figure \ref{fig:title_retrieval} shows the relative number of retrieved URIs when the title of the pages was queried.
Two observations can be made from this figure. The first is titles return more than $61\%$ top ranked URIs and another $4.5\%$ within
the top $10$. The second is a binary retrieval pattern meaning the URI is either discovered within the top $10$ results
(this time including the top result) or is it (by our definition) undiscovered.
Only $1.2\%$ of all URIs were returned within the top $100$ but beyond the top $10$.
\subsection{Similarity of Search Results}
Thinking about web pages' titles as search engine queries lets us intuitively identify three special cases of search results:
\begin{enumerate}
\item \textit{Aliases}, meaning two or more URIs point to virtually the same content where the URIs may or may not canonicalize to the same value 
\item \textit{Duplicates}, meaning two or more pages hold duplicated content or a large subset of the other
\item \textit{Title collisions}, meaning two or more pages share the same title but their content is very different.
\end{enumerate}
To further illustrate these special cases let us explore the following examples:
The wikipedia page for ``Lateef''\footnote{\texttt{http://en.wikipedia.org/wiki/Lateef}}
and ``Lateef The Truth Speaker''\footnote{\texttt{http://en.wikipedia.org/wiki/Lateef\_the\_Truth\_Speaker}}
are the same and hence can be considered as aliases.
The former has a note saying ``(Redirected from Lateef the Truth Speaker)'', but there is no HTTP notification about their equivalence.
Note that Google indexes the former, and Yahoo indexes the latter, however, neither search engine produces duplicate links.

%
\begin{figure*}[t]
\begin{center}
 \subfigure[$o$ for Discovered URIs]{\label{fig:to_rank_dis}\includegraphics[scale=0.35]{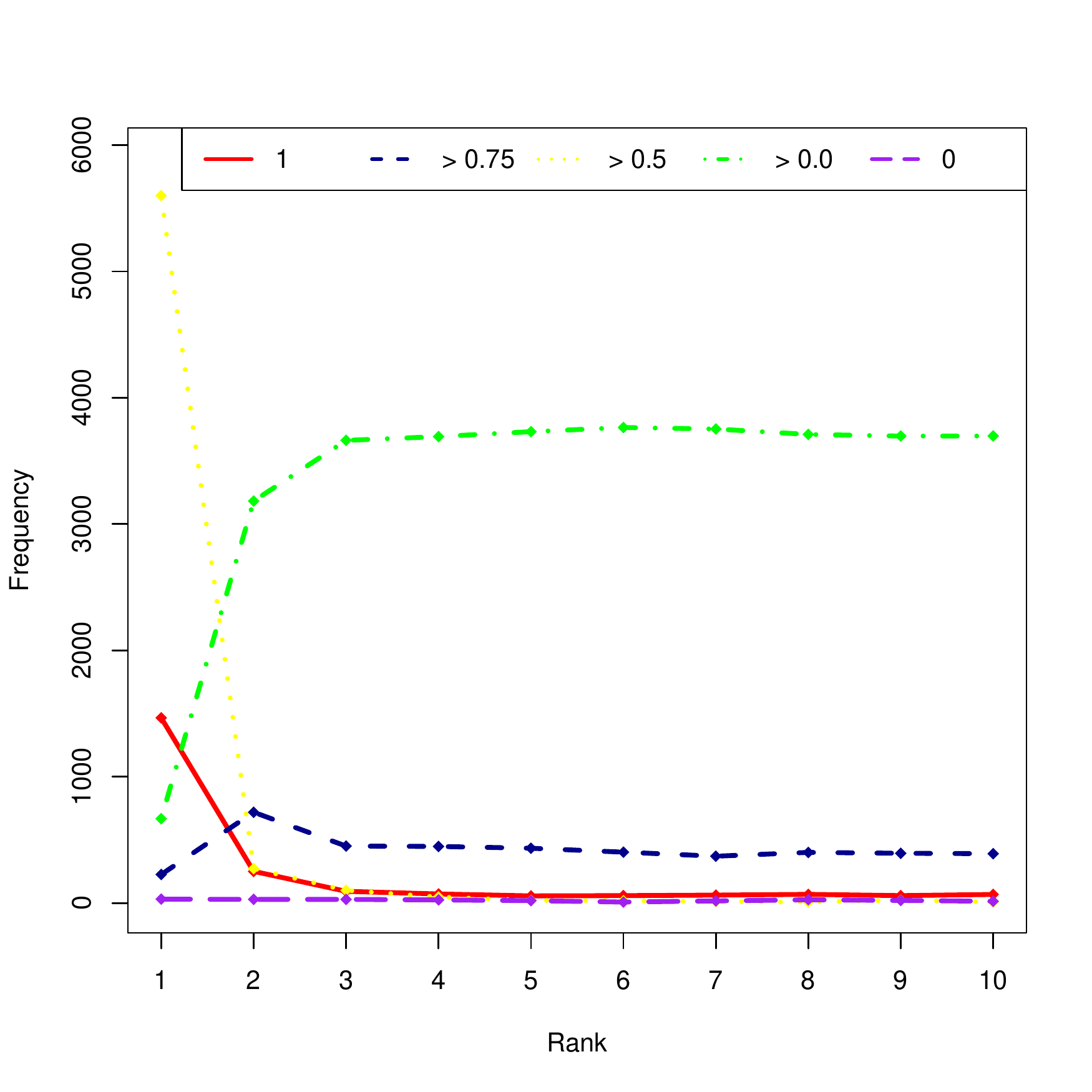}}
 \subfigure[$o$ for Undiscovered URIs]{\label{fig:to_rank_undis}\includegraphics[scale=0.35]{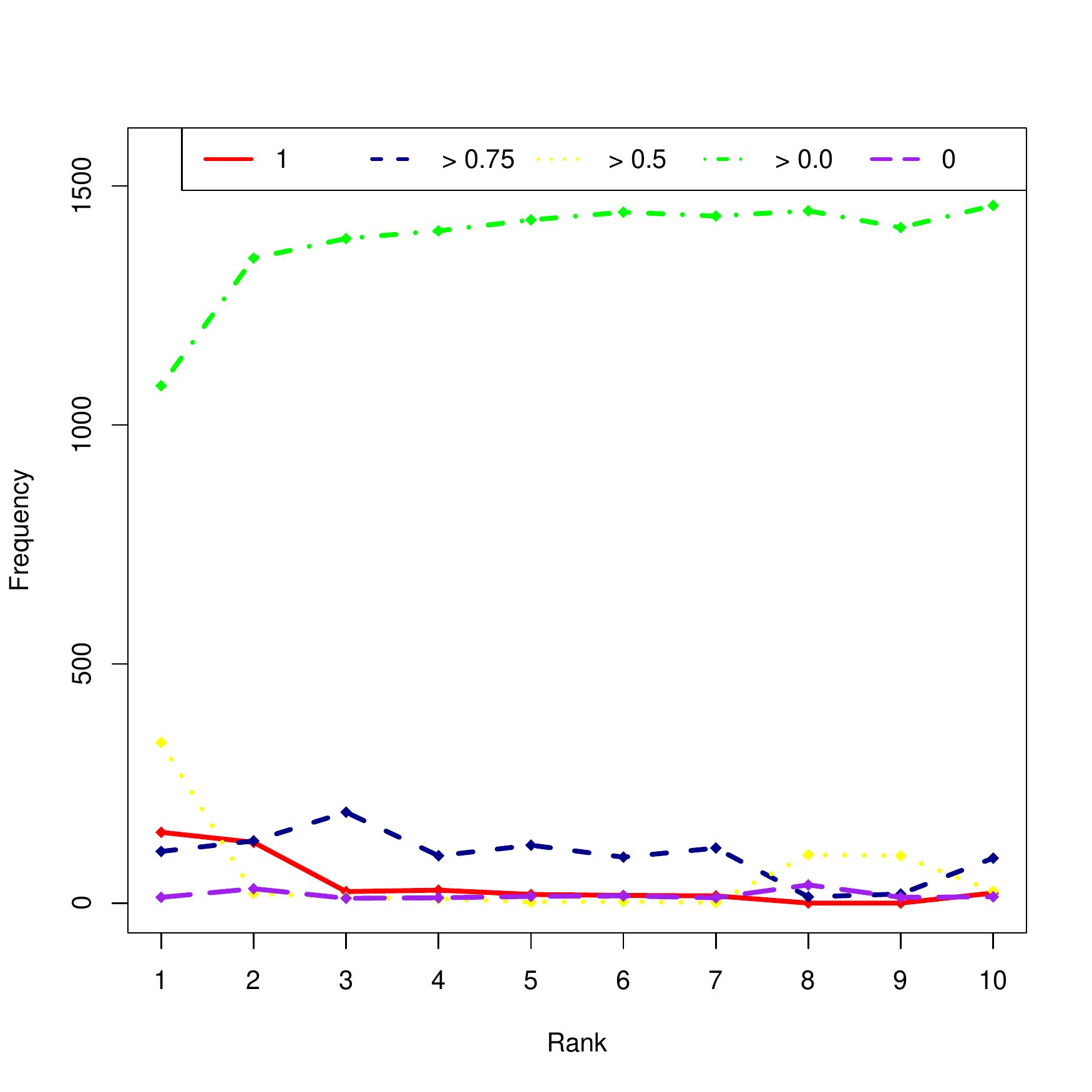}}
 \subfigure[$s$ for Discovered URIs]{\label{fig:shingle_rank_dis}\includegraphics[scale=0.35]{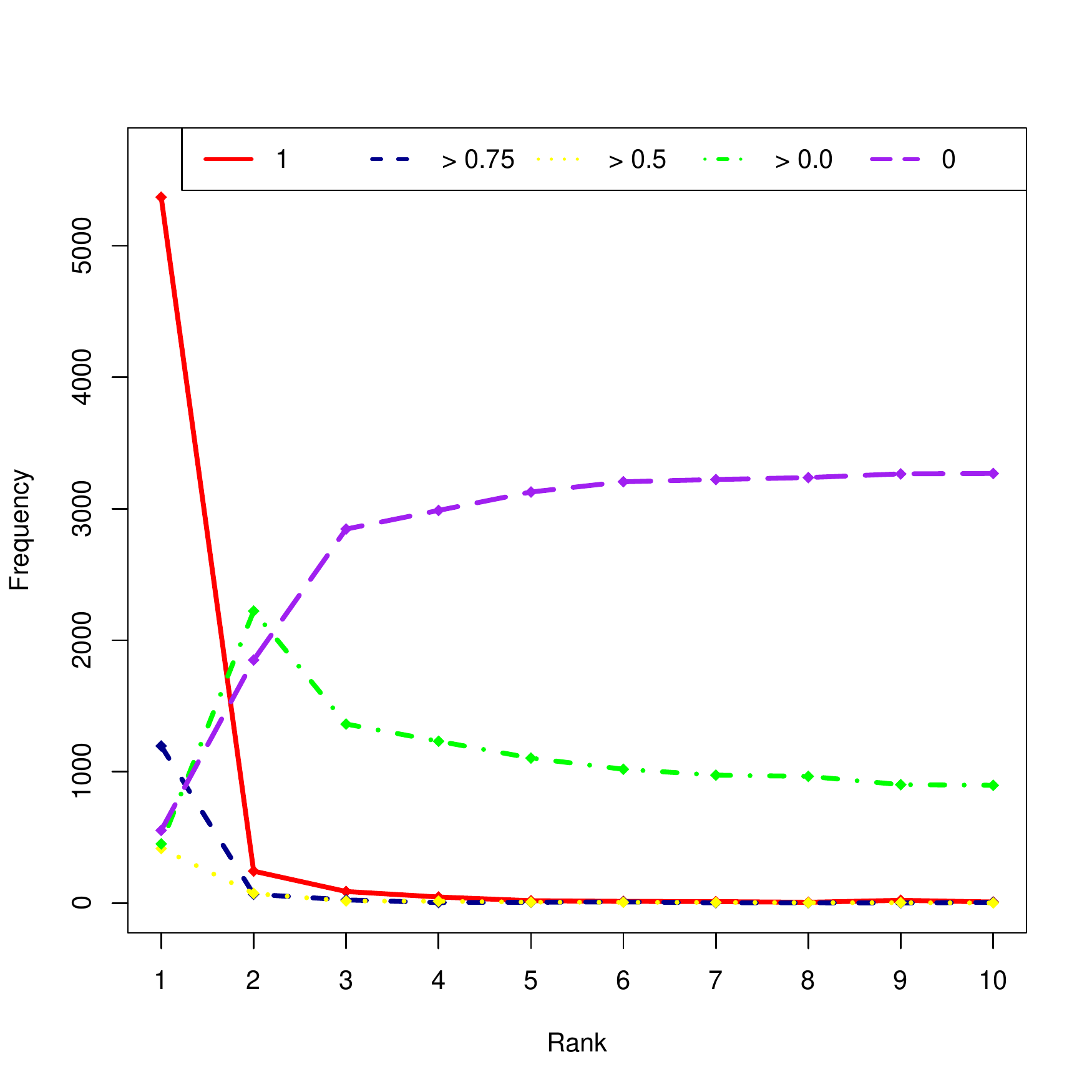}}
 \subfigure[$s$ for Undiscovered URIs]{\label{fig:shingle_rank_undis}\includegraphics[scale=0.35]{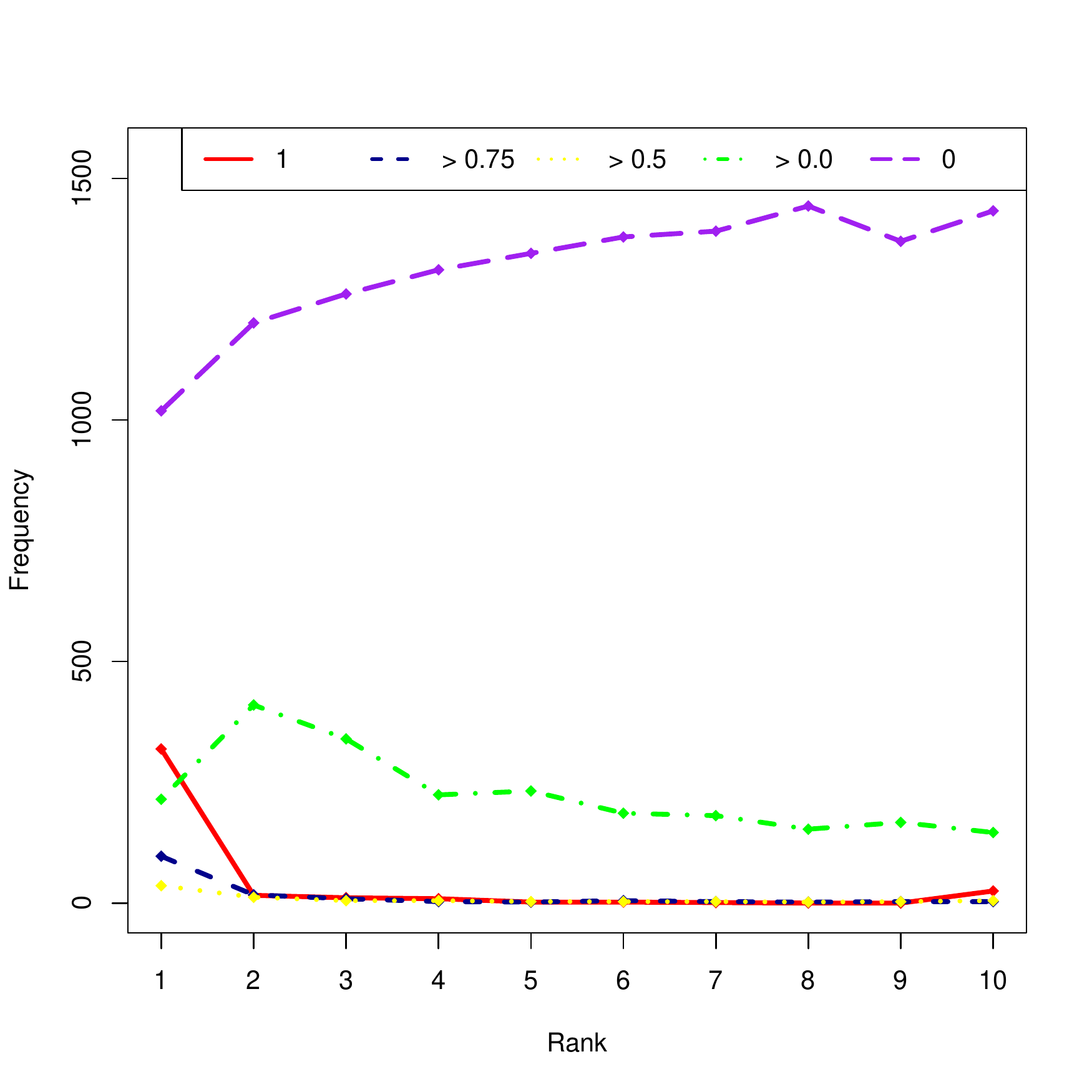}}
 \caption{Five Classes of Term Overlap ($o$) and Shingle Values ($s$) by Rank for Discovered and Undiscovered URIs. $o,s = 1$; $1 > o,s \ge 0.75$; $0.75 > o,s \ge 0.5$; $0.5 > o,s > 0.0$; $o,s = 0$}
 \label{fig:to_rank}
\end{center}
\end{figure*}
The wikipedia page for ``baseball balk''\footnote{\texttt{http://en.wikipedia.org/wiki/Balk}}
and the answers.com page\footnote{\texttt{http://www.answers.com/topic/balk}} can be considered duplicates. 
The two pages share a lot of overlap, in part because they both quote from the rule book. The answers.com page additionally includes
wikipedia content as well as other sources\footnote{\texttt{http://en.wikipedia.org/wiki/Answers.com}}.
Baeza-Yates et al. \cite{baeza-yates:genealogical_trees} explored the notion of genealogical trees for web pages where children would share
a lot of content with their parents and are likely to become parents themselves. Our notion of duplicates is similar to this concept.

An example for title collisions are the following five pages:
%
%
{\scriptsize http://www.globalrei.com/photos.php?property\_ID=70694}\\
{\scriptsize http://www.globalrei.com/photos.php?property\_ID=70694} \\
{\scriptsize http://www.globalrei.com/631-Westover-Aveune-a70694.html} \\
{\scriptsize http://www.globalrei.com/properties.php} \\
{\scriptsize http://www.globalrei.com/about.php} \\
{\scriptsize http://www.globalrei.com/globalrei/frm/3265/market\_information/} \\
All pages have the same title ``Welcome to my new website!'' but their content is very different. In fact the last URI even returns 
a customized $404$ error page.

In order to identify these three cases and reiterate our argument that web pages' titles perform well in search we investigate the 
similarity of the top $10$ results with the originating URI (the URI whose title was used as a query against the search API).
We used two methods to explore the similarity: normalized term overlap and w-shingles \cite{broder:shingles,broder:syntactic_clustering}
with a size of $w=5$.
We simplify the notion of retrieval to a binary scenario meaning either the URI was discovered or not. We define a URI as discovered if it
was returned within the top $10$ search results including the top rank. For all other cases the URI is considered undiscovered. We are aware that
we discriminate against URIs ranked between $11$ and $100$ with respect to our earlier measure of retrieval. However, the binary pattern 
which is obvious in Figure \ref{fig:title_retrieval} supports this simplification.

We display both, the overlap ($o$) and shingle values ($s$) divided in five classes.
Figure \ref{fig:to_rank_dis} displays the occurrence frequency of normalized term overlap values for all discovered URIs. 
The top rank is dominated by an overlap between $50\%$ and $75\%$. The fact that only $1466$ URIs located top ranked have the perfect overlap 
despite the more than $60\%$ top ranked URIs shown in Figure \ref{fig:title_retrieval} indicated that the content of the pages has changed between
the time we crawled the page and queried the search engine with its titles. From rank three on the most frequent overlap value is between $1\%$
and $50\%$.
Figure \ref{fig:to_rank_undis} shows a similar graph for all undiscovered URIs. 
The lower overlap class of values between
$1\%$ and $50\%$ throughout the ranks stands out. The perfect overlap 
is only noticeable for the top rank which indicates discovered aliases. 
The class with values between $50\%$ and $75\%$ occurs most frequently for the top rank as well which indicates
the discovery of duplicates.

A different and potentially better measure of document similarity is w-shingles. The shingle value of $1$ ($s=1$) is an indicator of strong similarity
(note that it does not guarantee identical content) and a null value indicating no similarity between the shingles.
Figure \ref{fig:shingle_rank_dis} shows five classes of shingle values by rank for discovered URIs.
We see the dominance of the top rank with shingle value $s=1$ which is not surprising considering the great amount of URIs discovered top ranked
(see Figure \ref{fig:title_retrieval}). However, this optimal shingle value is achieved more often than URIs discovered top ranked with indicates
that we discovered aliases and duplicates since for those cases we expect shingle values to be high. The zero value is rather low for the top rank,
increases for rank two and three and then levels off.
Figure \ref{fig:shingle_rank_undis} shows the same classes of shingle values for undiscovered URIs. As expected in this scenario the zero
value occurs very frequently. However, the $300$ occurrences of $s=1$ for the top rank is surprisingly good. It indicates that here as well
we have discovered a number of duplicates and aliases within the top ranks unlike the original URI.
\section{Title Evolution Over Time}
It is our intuition that web pages' titles change less frequently and less significantly than the web pages' content.
The title supposedly reflects the general topic of a page which naturally changes less often than its content.
With all from the IA provided copies downloaded we are able to investigate the temporal aspect of title changes. 
However, both, the time intervals in which the IA makes copies of web pages as well as the dates of the earliest and latest copies
of pages available in the IA can vary. In order to be able to investigate the title evolution over time we need to generalize the
available copies of each URI. We define $60$ days time windows in which we pick one copy per URI as representative for
the according time and URI. We define two such time windows per year, one in February and one in August starting with 2/2009.
Since the IA provides copies of web pages from as early as $1996$ we have a total of $27$ time windows and hence a maximum
of $27$ representative copies per URI from the last $14$ years.

\begin{figure}[!]
\centering
 \includegraphics[scale=0.5]{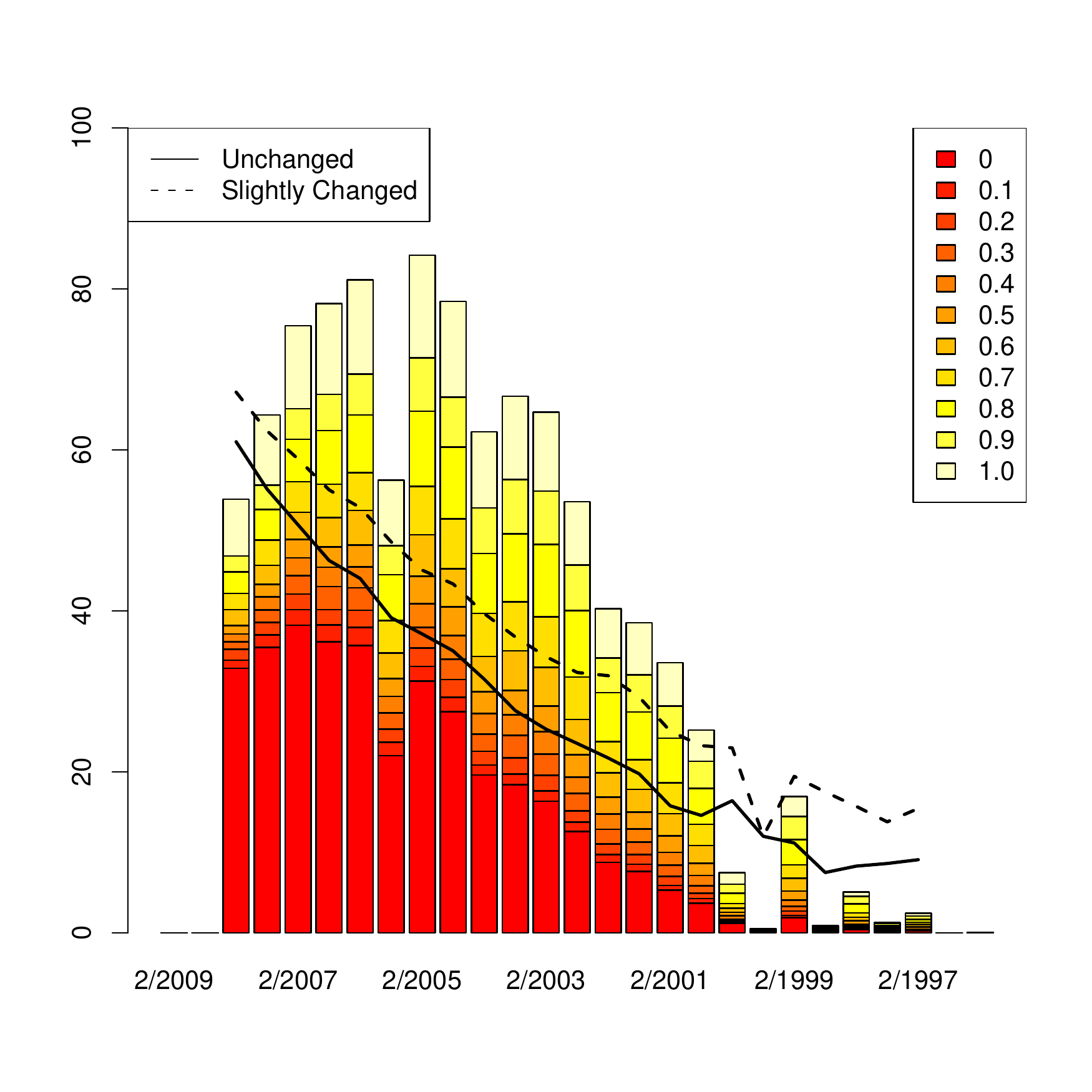}
 \caption{Title Edit Distance Frequencies per Internet Archive Observation Interval}
 \label{fig:title_ed}
\end{figure}

The Levenshtein edit distance gives a measure of how many operations are needed to transform on string into another. We compute
the edit distance between the titles obtained from the pages as they were downloaded in August of 2009 (our baseline)
and our representative copies from the IA.
We normalize the distance to a value between zero and one where one means the titles are completely dissimilar and a value of zero
indicates identical titles.
%
%
%
%
%
The edit distance distribution per time interval is shown in Figure \ref{fig:title_ed}.
The time intervals are represented on the x-axis with the most recent on the far left decreasing to the right. 
The number of available copies in the early years of the IA is rather sparse. There are no copies of our web pages
available in the first two time intervals (2/2009 and 8/2008). A possible explanation is the IA internal quarantine period
mentioned earlier.
However, for the third time interval we already find copies of roughly $55\%$ of all URIs.
The graph reveals that about half of the available titles from the more recent copies up until 2/2006 are identical or at least
very similar to the baseline.
We for example find copies of about $80\%$ of all URIs for time interval 2/2007 and more than half of those titles 
have an edit distance of zero.
This ratio drops for earlier copies. For the time intervals in 2002 for example we see only about $30\%$ of the available titles with
a distance value of zero. We have to keep in mind though that copies for only about $40\%$ of our URIs are availabe from that time.
From 2006 on it seems that the percentage of low edit distance values decreases while the amount of higher distances increases.
The solid line in Figure \ref{fig:title_ed} indicates the probability, based on our corpus, that a title of a certain age is
unchanged. Our data reveals that for copies as old as four years we have a $40\%$ chance of an unchanged title.
We define titles that have an edit distance value of $\le0.3$ as titles with only minor changes compared to the baseline.
The dashed line represents the chances for such titles given their age. We can see that for copies of web pages as old as $5.5$
years we have a probability of at least $40\%$ that the title has undergone only minor changes.
%
%
%
\begin{figure}[!]
\centering
 \includegraphics[scale=0.24]{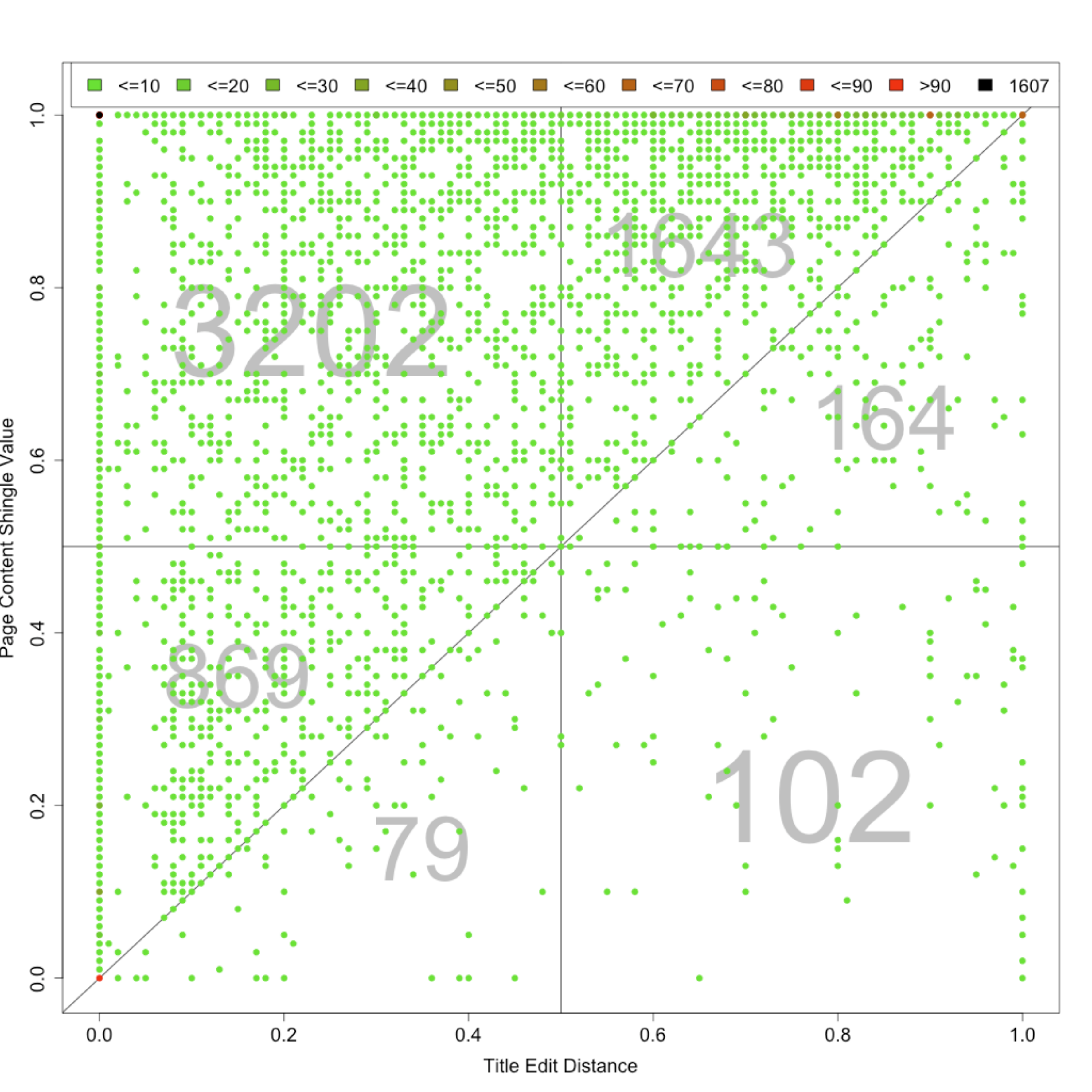}
 \caption{Title and Document Changes}
 \label{fig:ed_doc_change}
\end{figure}
%

If a page's title changes less frequently over time than its content the title could constitute a reliable search engine
query for discovering missing web pages.
To prove this intuition we computed shingle values for all available IA copies in the above mentioned time intervals and our baseline
version of the according page downloaded in August of $2009$. We normalized these values so that zero indicates a very similar page and
the value of one a very dissimilar page content. In order to compare these values with the edit distance of our titles in two dimensions
we computed the average of all available copies in our time intervals per URI.

Figure \ref{fig:ed_doc_change} shows the average normalized edit distance on the x-axis and the average normalized shingle value of the according
URI on the y-axis. Both values are rounded to the nearest tenth.
The color indicates the overlap per point or in other words the amount of times a certain point was plotted.
The palette starts with a basic green indicating a frequency of less or equal than $10$ and transitions into a solid red representing a frequency
of more than $90$.
The semi-transparent numbers represent the total amount of points in the according quarters and its halves.
The pattern is very apparent. The vast majority of the points are plotted with an average shingle value of above $0.5$ and an average edit
distance of below $0.5$. The most frequent point in fact is plotted more than $1,600$ times. It is colored black and located at the coordinates
$[0,1]$ meaning close to identical titles and very dissimilar content.
The point at $[0,0]$ is plotted $122$ times and hence somewhat significant as much as some points with a shingle value of one and an edit
distance of above $0.5$.

Figure \ref{fig:ed_doc_change} supports our intuition that titles change less significantly over time than the
pages' content. Therefore we claim titles to be the more robust technique with respect to content changes to discover missing web pages
compared to techniques such as LSs which are based on extracting the most salient terms from the pages content.
\begin{figure*}[!t]
\begin{center}
 \subfigure[All Titles]{\label{fig:title_length}\includegraphics[scale=0.27]{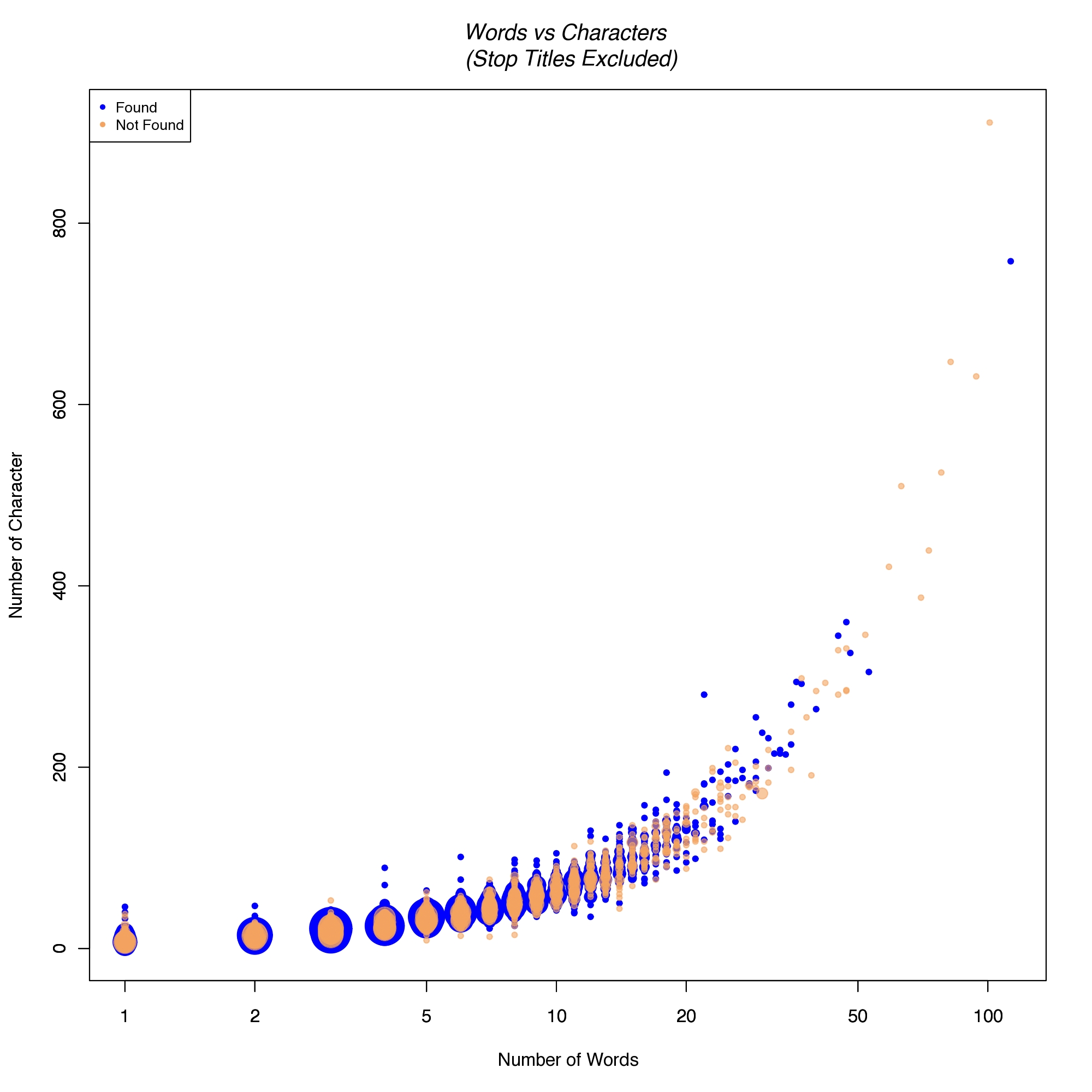}}
 \subfigure[Titles with Nine or Fewer Terms]{\label{fig:title_length_zoom}\includegraphics[scale=0.27]{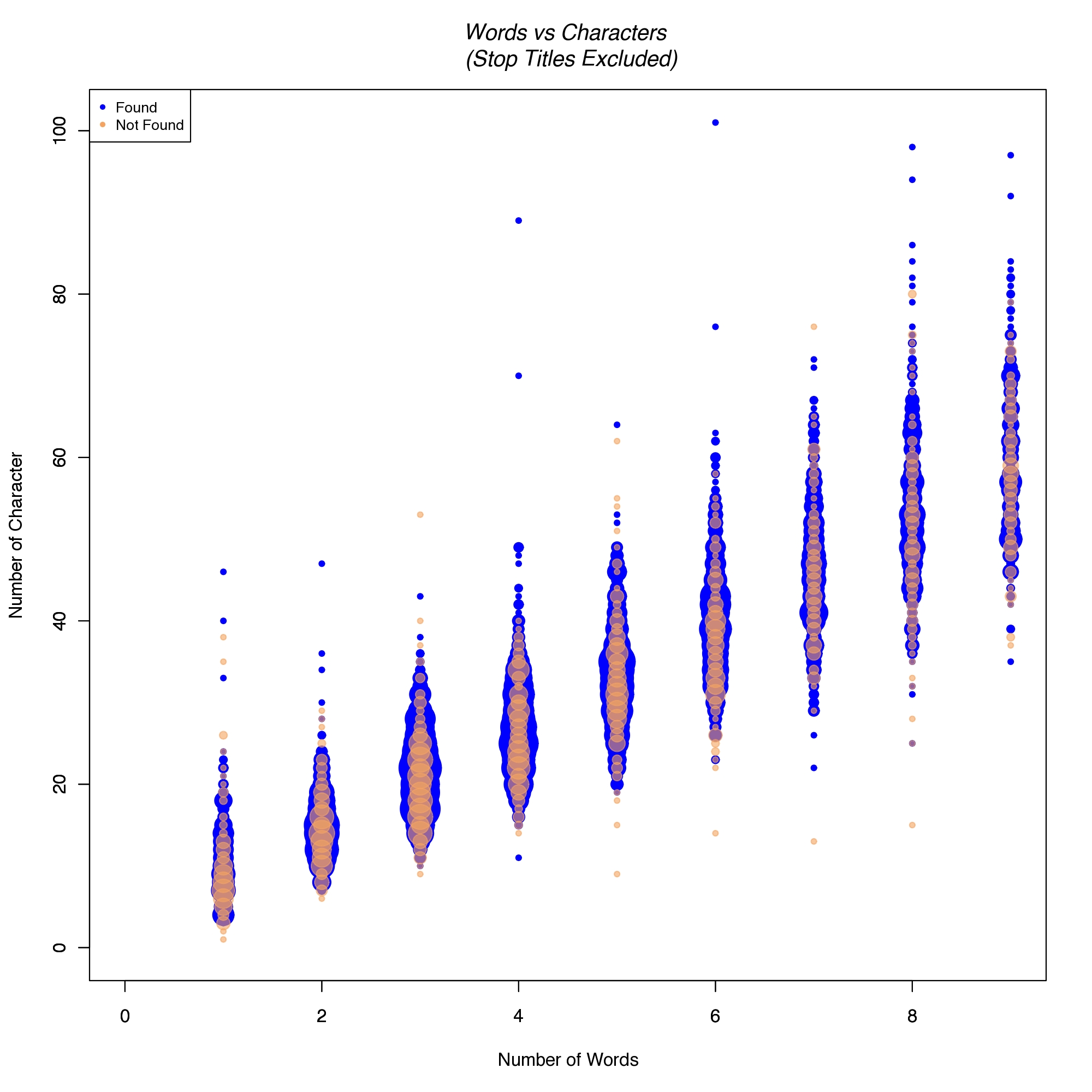}}
 \caption{Title Length in Number of Terms and Characters Distinguished by URI Found and Not Found}
 \label{fig:title_length_chars}
\end{center}
\end{figure*}
\section{Title Performance Prediction}
The example shown in Table \ref{tab:ls_title_example} supports the intuition that not all titles are equally good in terms of
their performance as search engine queries for the discovery of web pages.
We are interested in a method to analyze any given title in real time and give a probability for its usefulness for our search.
If we can identify a candidate bad title, we will not waste our time but proceed with the generation of a LS as
the more promising approach for web page discovery.
Ntoulas et al. \cite{ntoulas:detecting_spam} used methods based on web pages' content to identify spam web pages.
One of their experiments shows that web pages' titles consisting of more than $24$ terms are likely to be spam.
This result confirms that there are indicators by which we can predict the usefulness of titles for search.
%
Figure \ref{fig:title_length} displays the composition of our titles with respect to the number of terms (on the x-axis) and number of
characters (y-axis) they contain. The two different colors of the dots represent cases where URIs were found and not found.
This graph supports the findings of Ntoulas et al. since it is visible that titles containing between two and $10$ terms and less
than $200$ characters return more URIs discovered than undiscovered and hence can be considered good titles. 

The graph further shows that although search engines may enforce query limitations (number of query terms or number of characters)
on their APIs, we can not say with certainty that queries that exceed the limitations will be unsuccessful. That means in case the servers
silently truncate the queries the titles still may hold enough salient information in the first $n$ characters/terms that do not exceed the
possible limit to discover the URI.
Figure \ref{fig:title_length_zoom} shows the same information for titles with nine or less terms. These titles account for more than $70\%$ of
all titles in the entire corpus.

Most search engines automatically filter stop words (language dependent) from a query string since they are not
crucial when it comes to describing a pages' content. For titles, however, our intuition is that we can identify additional terms
that would not necessarily occur in a common stop word list for the English language but do not contribute to uncover the
``aboutness'' of a web page. 

We analyze our corpus of $6,875$ titles and identify those that (used as the query string) do not lead to rediscovering the
originating page. 
We call these titles \textbf{stop titles}.
For the sake of simplification we narrow the rediscovery to a binary value meaning URIs that are returned within the top
ten search results including the top rank are considered discovered and all remaining URIs are not. 
Some of the most frequent stop titles in our corpus are \textit{home}, \textit{index}, \textit{home page}, \textit{untitled document}
and \textit{welcome}.
We further argue that terms such as \textit{main page}, \textit{default page} and \textit{index html} should be added to the
list of stop titles even though they did not occur in our dataset.
The experienced Internet user and website creator will recognize most of these terms as the default title setting of various web page
generating tools.
\begin{figure*}[!t]
\begin{center}
 \subfigure[Stop Titles and Total Number of Terms]{\label{fig:word_ratio}\includegraphics[scale=0.35]{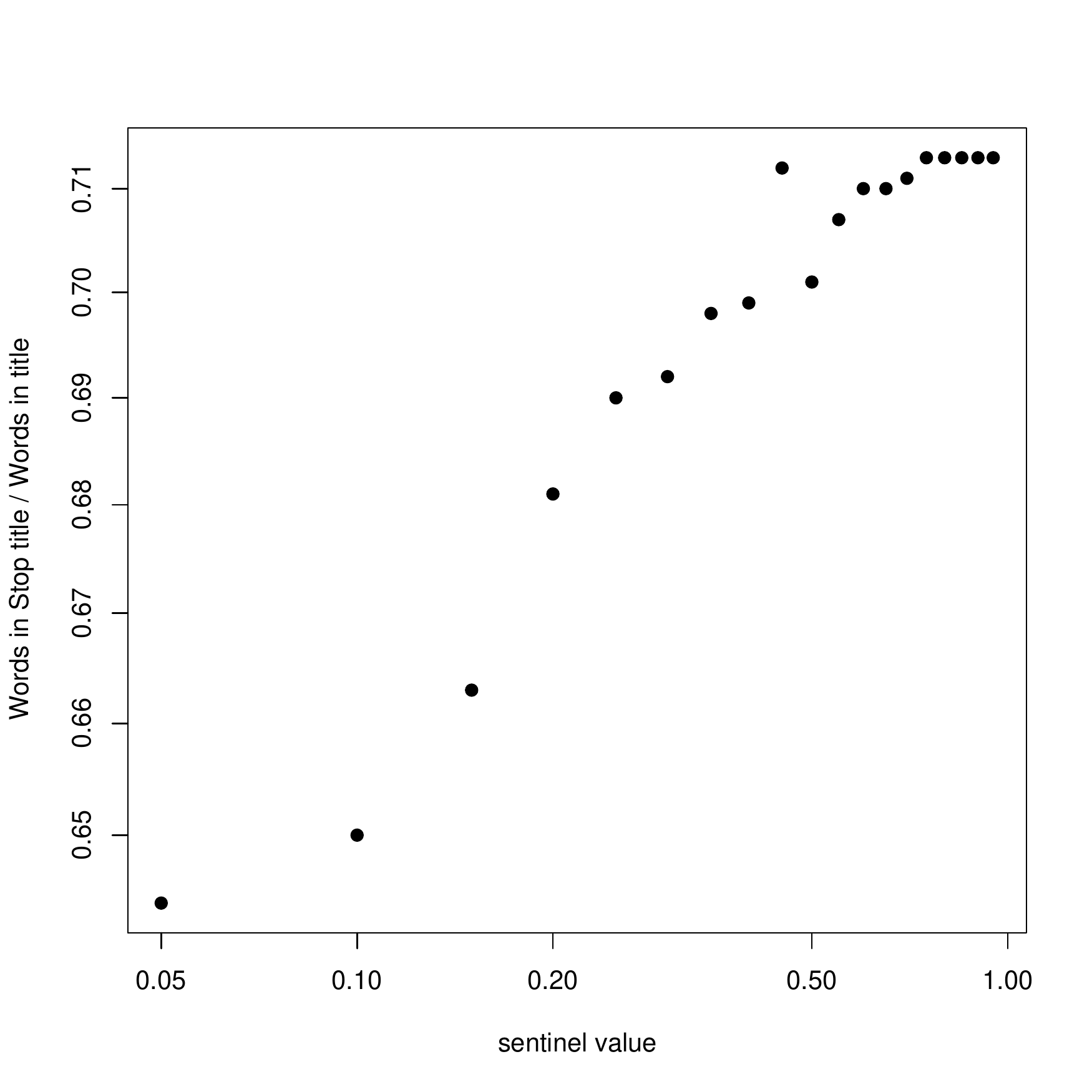}}
 \subfigure[Stop Title Characters and Total Number of Characters]{\label{fig:char_ratio}\includegraphics[scale=0.35]{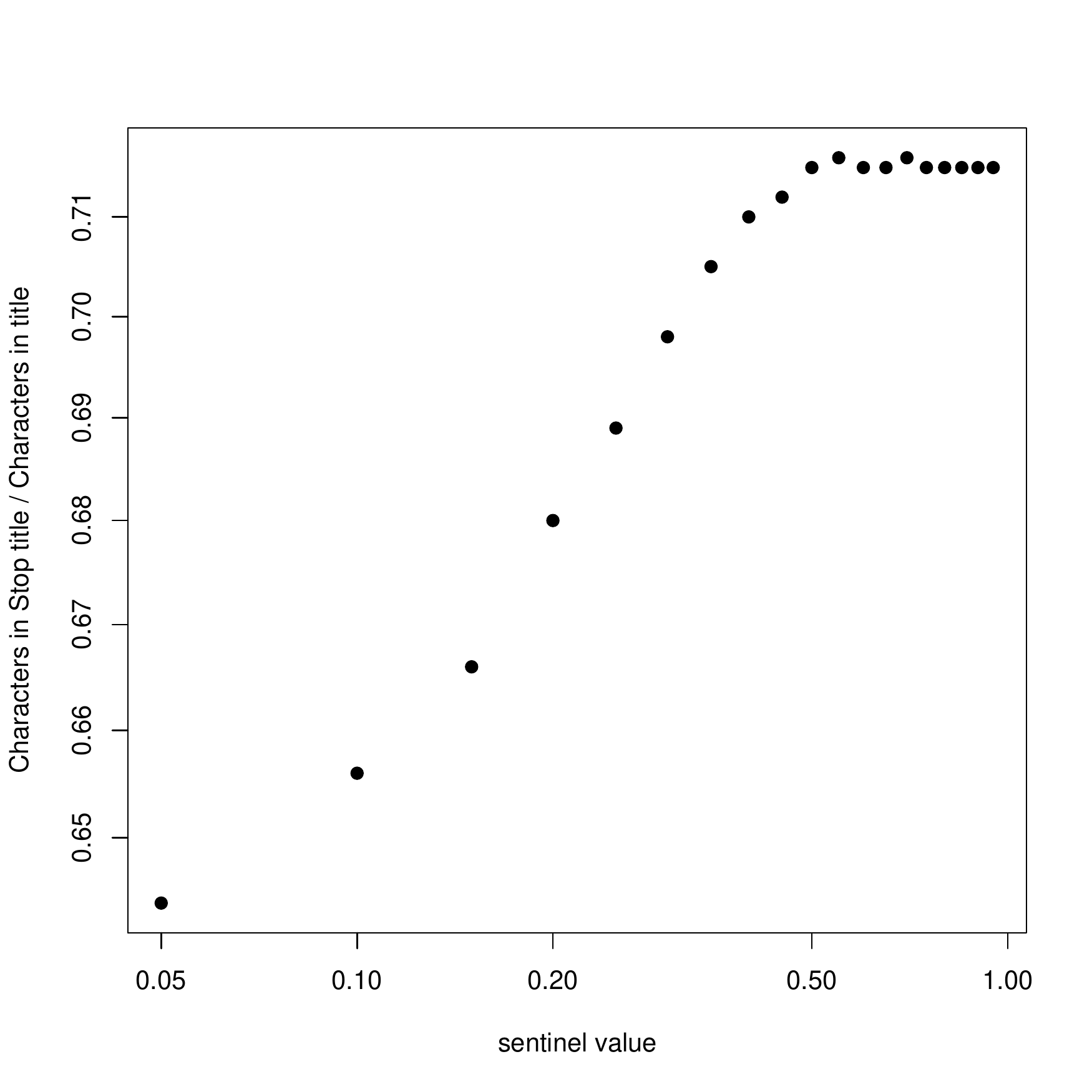}}
 \caption{Upper Bounds for Ratios of Number of Stop Titles in the Title and Total Number of Terms in the Title and Number of Stop Title Characters and Total Number of Characters in the Title}
 \label{fig:ratios}
\end{center}
\end{figure*}
With the list of stop titles our approach is to automatically identify bad titles. The trivial case is to match a given title
with all of the stop titles and if we find a match the title is classified as bad. The second approach is to compute the ratio of
stop titles and total number of terms in the title. The analysis on our corpus has shown that if this ratio is greater than $0.75$
the likelihood of the title performing poorly is very high and hence the title should be dismissed. Figure \ref{fig:word_ratio}
shows the sentinel value indicating the upper bound for the ratio based on our corpus and the binary discovered/undiscovered
classification of all titles.
\begin{table}
 \centering
  \begin{tabular}{|c|c|c|c|} \hline
	\multicolumn{2}{|c|}{}&\multicolumn{2}{c|}{\textbf{Actual}} \\ \cline{3-4}
	\multicolumn{2}{|c|}{}&\textbf{Found}&\textbf{Not Found} \\ \hline
	\multirow{2}{*}{\textbf{Predicted}}&\textbf{Found}&66\%&0.42\% \\ \cline{2-4}
	&\textbf{Not Found}&28.27\%&5.28\% \\ \hline
  \end{tabular}
  \caption{Confusion Matrix for Stop Titles / Total Number of Words}
 \label{tab:word_matrix}
\end{table}
Table \ref{tab:word_matrix} shows the confusion matrix for the second approach based on our experiments. We can see that with the evaluated upper
bound for the ratio ($0.75$) we obtain a total match of more than $71\%$ and hence a mismatch of $28\%$.

The third approach is based on number of single characters in the title.
That means if a title contains stop titles we determine the ratio of number of characters in the stop title and number of
total characters of the title.
Here too (as shown in Figure \ref{fig:char_ratio}) a ratio greater than $0.75$ predicts a poor performance.
%
%
Table \ref{tab:char_matrix} shows the according confusion matrix the third approach based on our experiments. The upper bound (also $0.75$)
accounts for similarly good numbers with a total match of more than $71\%$ but also achieves zero per cent false positives.
\begin{table}
 \centering
  \begin{tabular}{|c|c|c|c|} \hline
	\multicolumn{2}{|c|}{}&\multicolumn{2}{c|}{\textbf{Actual}} \\ \cline{3-4}
	\multicolumn{2}{|c|}{}&\textbf{Found}&\textbf{Not Found} \\ \hline
	\multirow{2}{*}{\textbf{Predicted}}&\textbf{Found}&66.45\%&0.0\% \\ \cline{2-4}
	&\textbf{Not Found}&28.48\%&5.07\% \\ \hline
  \end{tabular}
  \caption{Confusion Matrix for Number of Characters in Stop Titles / Total Number of Characters}
 \label{tab:char_matrix}
\end{table}
\section{Future Work} \label{sec:futwork}
We see several aspects of future work for this body of research.
So far we have conducted our experiments with randomly sampled pages from the Open Directory Project meaning we used
pages that are actually not missing. We are collaborating with several institutions such as the IA, the Library of Congress and the California
Digital Library in order to generate an extensive corpus of missing web pages. We will apply our methods against this corpus and be able to benchmark
their individual retrieval performance.

We have shown statistics of how many URIs have been re-discovered with titles and LSs but a thorough investigation of what (kind of) URIs were not
discovered may reveal valuable information that will help improve the entire retrieval process. For example we would expect pages about specific
and possibly abstract academic topics easier to rediscover than pages with very generic content that may even be used for search engine optimization
techniques. 

Besides titles and LSs we can think of various different methods to form potentially well performing search engine queries for our purpose.
Tags used to annotate web pages and LSs based on the page neighborhood (in- and out-links) for example have been shown to be useful for web
search. Using tags and neighborhood based LSs is the approach that has to be taken in case we can not locate any copies of the missing page
neither in search engines caches nor in the IA. The URI can be used to query for tags and in- and out-links.
LSs consist of unigrams only but phrases or n-grams are promising for discovering similar pages. The performance evaluation of these methods
remain for future work.
\section{Conclusions} \label{sec:concl}
We have investigated the performance of web pages' titles for the purpose of rediscovering missing web pages by utilizing the
web infrastructure. Titles are commonplace and we have shown that they perform similarly well compared to lexical signatures with respect
to the number of discovered URIs returned top ranked.
We did not address issues such as URI canonicalization in this work and hence we consider our retrieval results to be a lower bound.
We have further analyzed the top $10$ results for all queries and found that regardless of whether the URI was discovered, potential
URI aliases and duplicates were returned which also supports the argument that titles form well performing queries.
We have used copies of pages from the Internet Archive to further confirm our intuition that titles decay over time.
However, we have provided evidence that the content of web pages not only changes more quickly but also more significantly.
Hence we argue that titles can be a more robust method to rediscover missing pages.
%

We have shown that not all titles are ``good titles'', at least as determined by search suitability.
With a thorough analysis of the composition of all titles in our data set we have provided a guideline to automatically
identify titles predicted to perform poorly for our purpose. We have distilled a list of stop titles that indicate the
title's retrieval quality. A major benefit of this process is that it can be done in real time.
\section{Acknowledgement} \label{sec:concl}
This work is supported in part by the Library of Congress.
%
%
%

%
\end{document}